\documentclass[preprint2]{aastex}

\usepackage{color}
\usepackage{hyperref}
\usepackage{amsmath}

\shorttitle{Silicon-carbon molecules}
\shortauthors{Steglich et al.}

\begin{document}


\title{Electronic transitions of jet-cooled SiC$_2$, Si$_2$C$_n$ ($n=1-3$), Si$_3$C$_n$ ($n=1,2$), and SiC$_6$H$_4$ between 250 and 710 nm \\ \vspace{0.5cm}\small{\textit{published in Astrophys. J. 801 (2015) 119}}}


\author{M. Steglich, J. P. Maier}
\affil{Department of Chemistry, University of Basel, Klingelbergstrasse 80, CH-4056 Basel, Switzerland}
\email{m.steglich@web.de, j.p.maier@unibas.ch}



\begin{abstract}
Electronic transitions of the title molecules were measured between 250 and 710 nm using a mass-resolved 1+1' resonant two-photon ionization technique at a resolution of 0.1 nm. Calculations at the B3LYP/aug-cc-pVQZ level of theory support the analyses. Because of their spectral properties, SiC$_2$, linear Si$_2$C$_2$, Si$_3$C, and SiC$_6$H$_4$ are interesting target species for astronomical searches in the visible spectral region. Of special relevance is the Si--C$_2$--Si chain, which features a prominent band at 516.4 nm of a strong transition ($f=0.25$). This band and one from SiC$_6$H$_4$ at 445.3 nm were also investigated at higher resolution (0.002 nm).
\end{abstract}


\section{Introduction} \label{sec_intro}
Silicon and carbon are among the most abundant refractory elements in the universe. They play an important role in astrochemical processes and are present in more than 70\% of all inter- and circumstellar molecules. Already in the 1920s, Merrill and Sanford observed strong absorption bands in the 420--550 nm wavelength range of cool carbon stars \citep{merrill26,sanford26,sarre00}, which were later ascribed to an Si--C--C molecule by \citet{kleman56}. It took almost three decades before the gas phase study of \citet{michalopoulos84} revealed the correct triangular geometry of this molecule. More recently, the electronic emission \citep{lloydevans00} and rotational spectra \citep[e.g. ][]{thaddeus84,gottlieb89,mueller12} were detected in circumstellar regions. The Merrill-Sanford bands have even been found outside our galaxy \citep{morgan04}. Other molecules containing both silicon and carbon were identified by radio astronomy in circumstellar shells \citep{mccarthy03}. These are SiC \citep{cernicharo89}, linear SiNC and SiCN \citep{guelin00, guelin04}, rhomboidal SiC$_3$ \citep{apponi99b}, and linear SiC$_4$ \citep{ohishi89}. Spectroscopic detections in the laboratory preceded the astronomical observations \citep{apponi99a,mccarthy99,mccarthy01,kokkin11}.

\begin{figure}[t]\begin{center}
\epsscale{1.0} \plotone{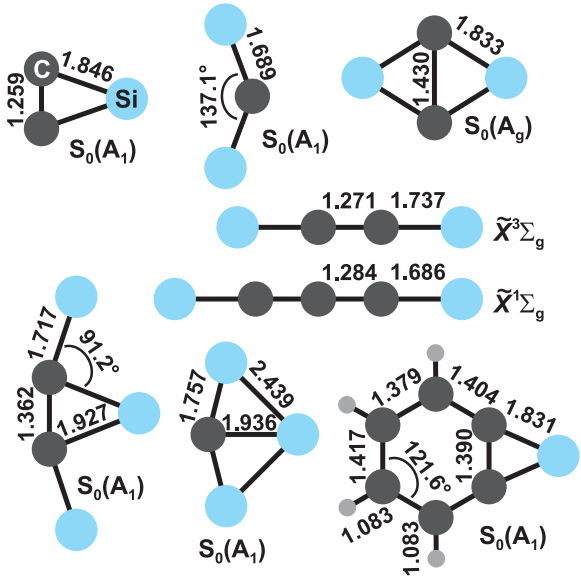} \caption{Calculated ground state structures of Si$_n$C$_m$ and SiC$_6$H$_4$, with bond lengths in \AA.} \label{fig_structures}
\end{center}\end{figure}

Several silicon-carbon clusters have been the subject of further laboratory investigations: vibrational modes of SiCH$_x$ ($x=3-6$), SiC$_2$H, SiC$_n$ ($n=2,4,5,7,9$), Si$_2$C$_n$ ($n=1-5$), and Si$_3$C$_n$ ($n=1,2$) could be identified by IR matrix isolation spectroscopy \citep{kafafi83,shepherd85,presillamarquez90,presillamarquez95,presillamarquez96,presillamarquez97,presillamarquez91,presillamarquez92,presillamarquez94,withey92,han97,ding99,ding00,kaiser05,le14a,le14b}. Gas phase IR spectra of SiC$_4$, Si$_2$C$_3$, and Si$_n$C$_m$ ($n+m=6$) have been obtained by \citet{orden94,orden95}, \citet{thorwirth11}, and \citet{savoca13}. Cations of SiC$_n$ ($n=3-69$) and anions of Si$_n$C$_m$ ($1\leq n \leq 7$, $1\leq m \leq 5$) have been characterized by gas phase ion mobility experiments \citep{fye97} and photoelectron spectroscopy \citep{nakajima95,duan02}, respectively. Photodissociation studies were performed on Si$_n$C$_m$ cations \citep[$n=3-10$, $m=1,2$;][]{ticknor05} and ionization energies were determined for SiC$_2$H$_x$ ($x=0-2$) \citep{kaiser12}. Electronic absorption spectra were measured mainly for small molecules in the past, such as SiC, SiCH, and SiC2 \citep{bernath88,brazier89,ebben91,butenhoff91b,grutter97,smith00,kleman56,weltner64,bondybey82,michalopoulos84,butenhoff91a}. Studies on larger systems were realized only recently. \citet{stanton05} analyzed the $\widetilde{C}$$^1\text{B}_1\leftarrow\widetilde{X}$$^1\text{A}_1$ transition of Si$_3$C, and \citet{kokkin14} investigated the optical spectra of silicon-terminated carbon chains, SiC$_n$H ($n=3-5$).

In light of several unassigned optical features of the interstellar medium, especially the diffuse interstellar absorptions bands (DIBs) on one hand, and an observed abundance of small silicon-carbon molecules in circumstellar environments on the other, a wider knowledge about the optical properties of such systems is of interest. The investigation at hand addresses this cause, presenting the UV-visible spectra of jet-cooled SiC$_2$, Si$_2$C$_n$ ($n=1-3$), Si$_3$C$_n$ ($n=1,2$), and SiC$_6$H$_4$ between 250 and 710 nm.

\section{Methods} \label{sec_exp}
Optical absorption spectra were measured indirectly by applying a resonant two-color two-photon ionization scheme (R2C2PI) and monitoring the ions (as a function of scanning laser wavelength) with a linear time-of-flight mass spectrometer (TOF-MS). Mixed silicon-carbon clusters were created by laser ablation (532 nm, 5 ns, $\approx$ 0.3 J/mm$^2$) of a crystalline Si rod in an He atmosphere containing 2\% acetylene (C$_2$H$_2$). In later experiments, SiC$_6$H$_4$ was also produced by discharging a gas mixture of He and phenylsilane (SiC$_6$H$_8$), the latter being vaporized in a bubbler. The molecular beam source contains a pulsed valve (5 bar backing pressure) creating a supersonic jet, which was skimmed 40 mm downstream to generate a collimated beam. A +300 V potential was applied to the skimmer to remove ions before neutrals entered the ionization region of the TOF-MS. The spectral scans were realized by counter-propagating the radiation of the laser into the molecular beam. Broad range scans between 250 and 710 nm were conducted with an optical parametric laser (OPL; 5$-$10 ns, 20 Hz, 0.1 nm bandwidth). A dye laser ($\approx$10 ns, 10 Hz, 0.002 nm bandwidth) was used for higher-resolution scans. All spectra were corrected for wavelength-dependent laser power variations. Molecules were ionized by the output of a second OPL (wavelength fixed at 215 nm), an ArF excimer laser (193 nm), or an F$_2$ excimer laser (157 nm), crossing the molecular beam perpendicularly. The jitter of both excimer lasers was about 40 ns, those of the other lasers 5$-$10 ns. Absorption bands with shorter excited state lifetimes are difficult to observe in R2C2PI using two different laser sources.

Quantum-chemical calculations were conducted to assist the spectral analysis. Geometries and vibrations of ground and excited states as well as vertical transition and ionization energies were predicted with density functional theory (DFT) and time-dependent DFT (TDDFT) as implemented in the Gaussian09 software package \citep{frisch13}. The B3LYP functional \citep{becke88,lee88} was applied in conjunction with the aug-cc-pVQZ basis set \citep{dunning89,woon93}. Vibrational progressions and rotational profiles were simulated with Gaussian09 and PGOPHER \citep{western10}.

\begin{figure*}\begin{center}
\epsscale{1.9} \plotone{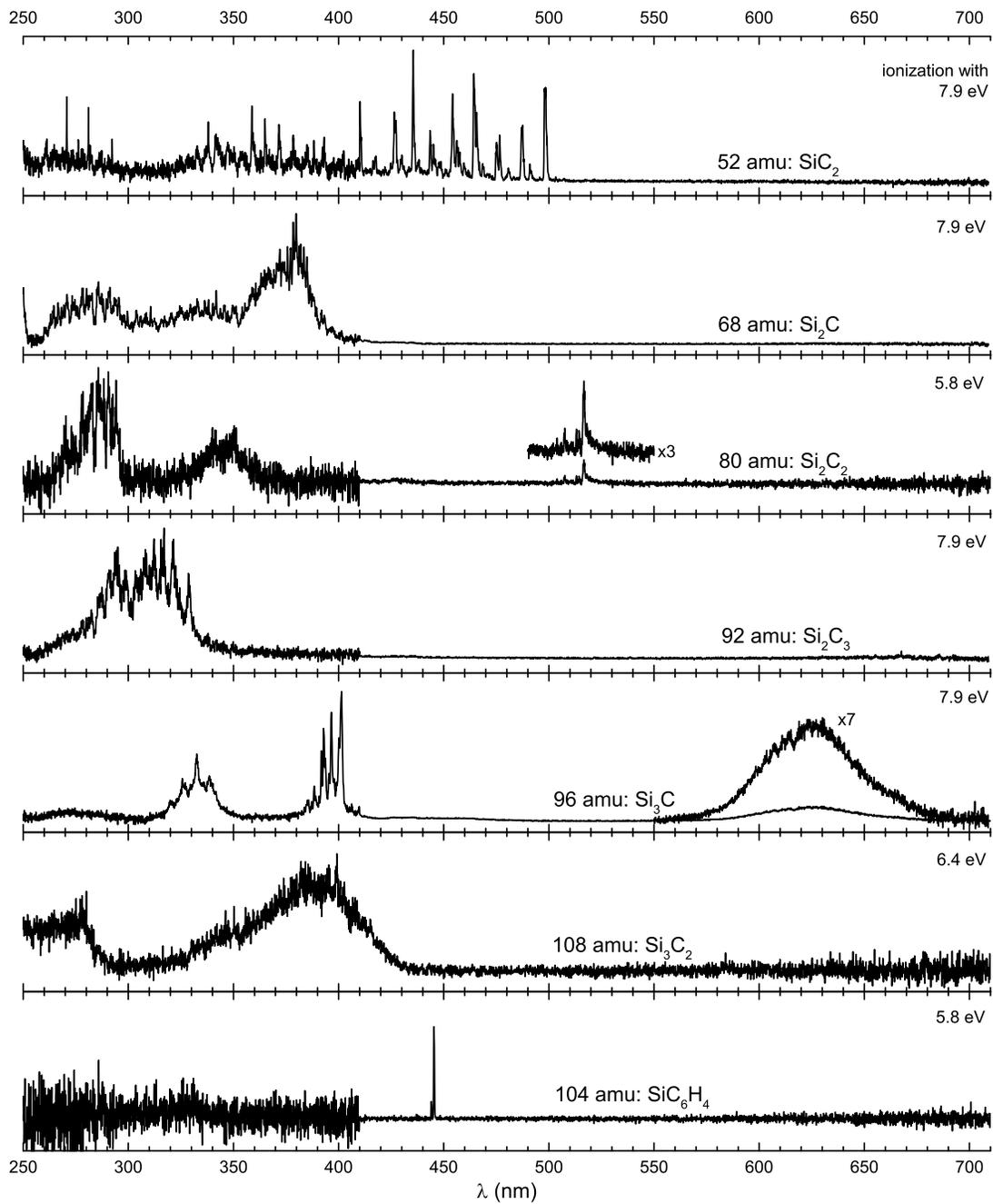} \caption{Low-resolution (0.1 nm) R2C2PI spectra. The energies of the ionizing photons are indicated.} \label{fig_overview}
\end{center}\end{figure*}

\begin{table*}
 \caption{Observed and calculated transitions and ionization energies.}
\begin{tabular}{|l|c|ccccc|} \hline
     			  	& observed origin		& 					&\multicolumn{2}{c}{theory (B3LYP/aug-cc-pVQZ)}		           &       					&                               	\\
     molecule  	& band (nm)			& symmetry		& assignment 									& position (nm)$^{\text{a}}$	& $f$        	& IE (eV)$^{\text{b}}$ 	\\ \hline \hline
SiC$_2$          	& 498.1	     		    	& C$_{2v}$		& S$_1$(B$_2$)$\leftarrow$S$_0$(A$_1$)	&       458 (485)    	&  0.018         			&       9.88 (9.73)            	\\ 
                     	& 393.0    			    	&           			& S$_2$(B$_1$)$\leftarrow$S$_0$(A$_1$)	&       368            	&  0.009       	  		&                                	\\
                     	& 292.2    			    	&           			& S$_5$(B$_1$)$\leftarrow$S$_0$(A$_1$)	&       253            	&  0.036         			&                                	\\ \hline
Si$_2$C          	& 379.7				& C$_{2v}$    		& S$_2$(B$_1$)$\leftarrow$S$_0$(A$_1$)	&       379 (451)	&  0.033         			&       9.19		            	\\
		          	& 						& 			    		& S$_3$(B$_2$)$\leftarrow$S$_0$(A$_1$)	&       373			&$4\cdot10^{-4}$		&      		                    	\\
  		        	& 350$^{\text{c}}$	& 			    		& S$_5$(A$_1$)$\leftarrow$S$_0$(A$_1$)	&       344            	&  0.023         			&       		                    	\\
		          	& 300$^{\text{c}}$	&			   		& S$_6$(B$_1$)$\leftarrow$S$_0$(A$_1$)	&       270           	&  0.007         			&       		                    	\\
		          	& 						&			   		& S$_7$(B$_2$)$\leftarrow$S$_0$(A$_1$)	&       266           	&  0.072         			&       		                    	\\
		          	& 						&			   		& S$_8$(A$_1$)$\leftarrow$S$_0$(A$_1$)	&       253           	&  0.002         			&       		                    	\\ \hline
Si$_2$C$_2$   	& 516.4                     	& D$_{\infty h}$   	& $\widetilde{C}^3\Sigma_u\leftarrow \widetilde{X}^3\Sigma_g$	&	465 (486)		&  0.254				&       7.58 (7.48)              \\
				&						& D$_{2h}$   	  	& S$_1$(B$_{2g}$)$\leftarrow$S$_0$(A$_g$)	& 	    492			&  0.0$^{\text{d}}$	&       9.01 (8.60)           	\\
			   	& 		                     	& D$_{2h}$   	  	& S$_2$(B$_{1g}$)$\leftarrow$S$_0$(A$_g$)	& 	    468			&  0.0$^{\text{d}}$	&       		                    	\\
			   	& 365$^{\text{c}}$	& D$_{2h}$   	  	& S$_3$(B$_{3u}$)$\leftarrow$S$_0$(A$_g$)	& 	    360			&  0.049				&       		                    	\\
			   	& 		                     	& D$_{\infty h}$   	& $\widetilde{E}^3\Pi_u\leftarrow \widetilde{X}^3\Sigma_g$			&	    385			&  0.015				&       		                    	\\
			   	& 295$^{\text{c}}$	& D$_{2h}$   	  	& S$_4$(B$_{3u}$)$\leftarrow$S$_0$(A$_g$)	& 	    281			&  0.013				&       		                    	\\
			   	& 						& D$_{2h}$   	  	& S$_8$(B$_{2u}$)$\leftarrow$S$_0$(A$_g$)	& 	    253			&  0.004				&       		                    	\\ 
			   	& 						& D$_{2h}$   	  	& S$_9$(B$_{1u}$)$\leftarrow$S$_0$(A$_g$)	& 	    245			&  0.023				&       		                    	\\
			   	& 		                     	& D$_{\infty h}$   	& $\widetilde{H}^3\Pi_u$$\leftarrow \widetilde{X}^3\Sigma_g$		& 	    291			&  0.040				&       		                    	\\
			   	& 		                     	& D$_{\infty h}$   	& $\widetilde{I}^3\Sigma_u\leftarrow \widetilde{X}^3\Sigma_g$	& 	    280			&  0.830				&       		                    	\\ \hline
Si$_2$C$_3$   	& 328.6				& D$_{\infty h}$	& $\widetilde{F}^1\Sigma_u\leftarrow \widetilde{X}^1\Sigma_g$	&      292 (305)	&  1.766           		&       8.22                    	\\
			   	& 298.9				& 					& $\widetilde{G}^1\Pi_u\leftarrow \widetilde{X}^1\Sigma_g$		&       279     		&  0.105           		&       		                    	\\ \hline
Si$_3$C          	& 680$^{\text{c}}$	& C$_{2v}$        	& S$_1$(B$_1$)$\leftarrow$S$_0$(A$_1$)	&       663     		&  0.002           		&       8.12 (7.77)             	\\
		          	& 401.3				& 			        	& S$_4$(B$_1$)$\leftarrow$S$_0$(A$_1$)	&       401 (404)	&  0.028           		&       		                    	\\
		          	& 338.5				& 			        	& S$_5$(B$_1$)$\leftarrow$S$_0$(A$_1$)	&       330     		&  0.017           		&       		                    	\\ 
		          	& 						& 			        	& S$_6$(B$_2$)$\leftarrow$S$_0$(A$_1$)	&       320     		&  0.019           		&       		                    	\\ \hline
Si$_3$C$_2$   	& 430$^{\text{c}}$	& C$_{2v}$        	& S$_3$(B$_2$)$\leftarrow$S$_0$(A$_1$)	&       404     		&  0.122           		&       7.43 (7.26)             	\\
		          	& 353$^{\text{c}}$	& 			        	& S$_5$(B$_1$)$\leftarrow$S$_0$(A$_1$)	&       343     		&  0.005           		&       		                    	\\
		          	& 290$^{\text{c}}$	& 			        	& S$_8$(B$_1$)$\leftarrow$S$_0$(A$_1$)	&       285     		&  0.005           		&       		                    	\\
		          	&						& 			        	& S$_9$(A$_1$)$\leftarrow$S$_0$(A$_1$)	&       275     		&  0.019           		&       		                    	\\ \hline
SiC$_6$H$_4$	& 445.3                     	& C$_{2v}$        	& S$_1$(B$_1$)$\leftarrow$S$_0$(A$_1$)	&       410 (465)	&  0.012          		&8.47 (8.24)$^{\text{e}}$	\\ \hline
\end{tabular}\\
\begin{tabular}{l}
\footnotesize $^a$vertical excitation energy (adiabatic value in brackets if calculated) \\
\footnotesize $^b$vertical ionization energy (adiabatic in brackets) \\
\footnotesize $^c$onset of broad absorption profile \\
\footnotesize $^d$unlikely assignment (see the text) \\
\footnotesize $^e$observed adiabatic IP = $(8.40 \pm 0.02)$ eV \\
\end{tabular}
\label{tab_results}
\end{table*}

\section{Results \& Discussion} \label{sec_results}
The theoretical ground state geometries of the observed molecules are shown in Fig. \ref{fig_structures}. The lowest Si$_n$C$_m$ isomers were recently predicted with the same method of theory \citep[among others;][]{duan10}. Several more species appeared in the mass spectra of the laser ablation products, including Si$_n$ up to $n=4$ and hydrogenated clusters, such as Si$_n$C$_m$H ($n=2,3$, $m=1,2$), but no (reproducible) absorption spectra could be measured under the applied experimental conditions. Initially, due to the same mass of 104 amu, the benzene-like SiC$_6$H$_4$ was deemed to be the linear molecule Si$_2$C$_4$. Calculations and further experiments, discussed in Sect. \ref{sec_SiC6H4}, led to the more conclusive assignment. 

Fig. \ref{fig_overview} displays the low-resolution spectra between 250 and 710 nm. Spectral assignments to electronic transitions are summarized in Table \ref{tab_results}. A discussion follows in the subsections below. 

\subsection{SiC$_2$} \label{sec_SiC2}
\begin{figure}[t]\begin{center}
\epsscale{1.0} \plotone{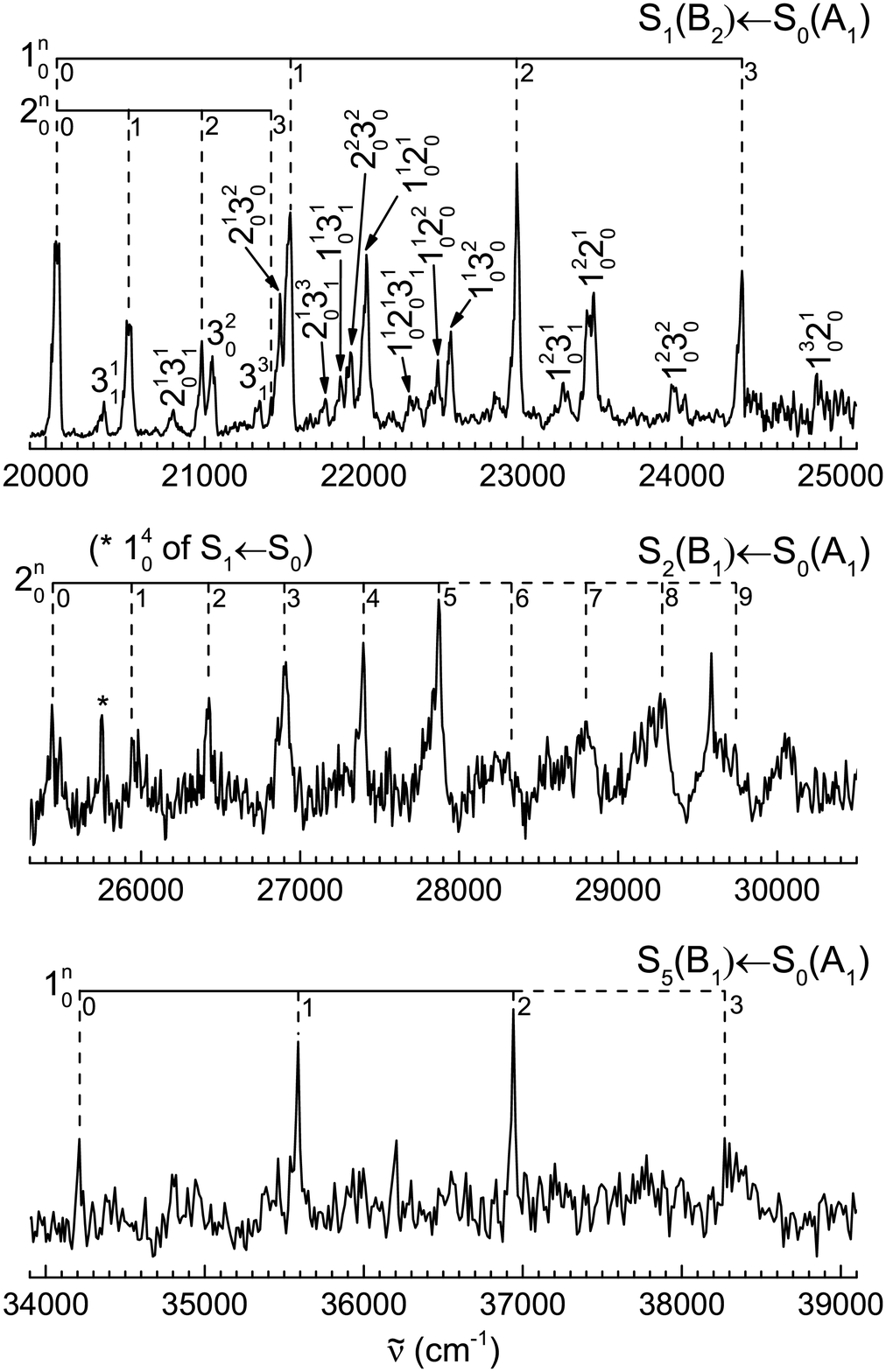} \caption{Vibrational assignments in the low-resolution (0.1 nm) spectrum of SiC$_2$.} \label{fig_SiC2}
\end{center}\end{figure}

After its first detection in the laboratory, SiC$_2$ was assumed to be linear by analogy with C$_3$ \citep{kleman56,bondybey82} until the R2C2PI study of the S$_1$(B$_2$)$\leftarrow$S$_0$(A$_1$) transition\footnote{or $\widetilde{A}$$^1$B$_2$$\leftarrow$$\widetilde{X}$$^1$A$_1$} by \citet{michalopoulos84} revealed a triangular ground state geometry. The S$_1$$\leftarrow$S$_0$ transition is the only electronic one that has been studied between 410 and 550 nm in the laboratory and in carbon stars \citep{butenhoff91a,sarre00}. The measurements presented here extend further into the UV, probing two additional transitions, assigned as S$_2$(B$_1$)$\leftarrow$S$_0$(A$_1$) with an apparent origin band at 393.0 nm and S$_5$(B$_1$)$\leftarrow$S$_0$(A$_1$) with an origin band at 292.2 nm. Vibrational assignments are denoted in Fig. \ref{fig_SiC2} and given in Table \ref{tab_SiC2}. Two singlet states (S$_3$(A$_2$) and S$_4$(B$_2$)) are predicted in between. The corresponding transitions from the ground state have oscillator strengths equal or close to zero.

\begin{table}
 \caption{Bands of SiC$_2$ (see Fig. \ref{fig_SiC2}).}
\begin{tabular}{|lccc|} \hline
     			  					& 						& $\lambda$ 			& $\Delta \nu$		\\
     		  						& band					& (nm)$^{\text{a}}$	& (cm$^{-1}$)	$^{\text{b}}$	\\ \hline \hline
S$_1$(B$_2$)					& 0$^0_0$				& 498.1				& 0					\\
								& 3$^1_1$				& 490.9				& 299				\\
								& 2$^1_0$				& 487.2				& 455				\\
								& 2$^1_0$3$^1_1$	& 480.6				& 737				\\
								& 2$^2_0$				& 476.5				& 916				\\
								& 3$^2_0$				& 475.0				& 982				\\
								& 3$^3_1$				& 468.3				& 1265				\\
								& 2$^3_0$				& 466.8				& 1352				\\
								& 2$^1_0$3$^2_0$	& 465.6				& 1407				\\
								& 1$^1_0$				& 464.2				& 1472				\\
								& 2$^1_0$3$^3_1$	& 459.5				& 1693				\\
								& 1$^1_0$3$^1_1$	& 457.5				& 1787				\\
								& 2$^2_0$3$^2_0$	& 456.2				& 1838				\\
								& 1$^1_0$2$^1_0$	& 454.1				& 1951				\\
								& 1$^1_0$2$^1_0$3$^1_1$	& 448.6		& 2247				\\
								& 1$^1_0$2$^2_0$	& 445.0				& 2378				\\
								& 1$^1_0$3$^2_0$	& 443.4				& 2482				\\
								& 1$^2_0$				& 435.4				& 2896				\\
								& 1$^2_0$3$^1_1$	& 429.9				& 3201				\\
								& 1$^2_0$2$^1_0$	& 426.4				& 3359				\\
								& 1$^2_0$3$^2_0$	& 417.7				& 3885				\\
								& 1$^3_0$				& 410.1				& 4313				\\
								& 1$^3_0$2$^1_0$	& 402.4				& 4782				\\
								& 1$^4_0$				& 388.2				& 5686				\\  \hline
S$_2$(B$_1$)					& 0$^0_0$				& 393.0				& 0					\\
								& 2$^1_0$				& 385.3				& 508				\\
								& 2$^2_0$				& 378.3				& 982				\\
								& 2$^3_0$				& 371.6				& 1465				\\
								& 2$^4_0$				& 364.9				& 1959				\\
								& 2$^5_0$				& 358.7				& 2432				\\  \hline
S$_5$(B$_1$)					& 0$^0_0$				& 292.2				& 0					\\
								& 1$^1_0$				& 280.9				& 1376				\\
								& 1$^2_0$				& 270.6				& 2730				\\  \hline

\end{tabular}\\
\begin{tabular}{l}
\footnotesize $^{\text{a}}$band maxima ($\pm$ 0.1 nm) \\
\footnotesize $^{\text{b}}$approx. band centers \\
\end{tabular}
\label{tab_SiC2}
\end{table}

The three normal modes of SiC$_2$ correspond to a C--C stretching vibration, $\nu_1$(a$_1$) = 1746.0 cm$^{-1}$ in the ground state, a stretch between the Si atom and the C$_2$ moiety, $\nu_2$(a$_1$) = 840.6 cm$^{-1}$, and an antisymmetric bending-type vibration, $\nu_3$(b$_2$) = 196.4 cm$^{-1}$ \citep{butenhoff91a}. The S$_1$$\leftarrow$S$_0$ spectrum features excitations to all three vibrational modes. The assignments in Table \ref{tab_SiC2} follow those of \citet{butenhoff91a} and \citet{sarre00}. In addition, the 1$^2_0$3$^2_0$, 1$^3_0$2$^1_0$, and 1$^4_0$ bands have been identified.

The vibrational pattern of the S$_2$$\leftarrow$S$_0$ system is dominated by almost evenly spaced bands which start at 393.0 nm and have a separation of 508 cm$^{-1}$ (between the first and second bands) to 473 cm$^{-1}$ (between fourth and fifth). This progression is attributed to the excitation of $\nu_2$. Since 2$^5_0$ is the most intense band, a substantial displacement between the Si atom and C$_2$ unit is suggested. After $n=5$, bands in the 2$^n_0$ series become broad and are hardly discernible.

Three narrow bands separated by 1372 and 1356 cm$^{-1}$ are observed in the S$_5$$\leftarrow$S$_0$ transition. The first one, presumably the origin band, appears at 292.2 nm. A weak and broad fourth band seems to follow the progression at the edge of the scanning range where the laser power was quite low. The spacing between the bands is only 100 cm$^{-1}$ lower than the energy of $\nu_1$ in the S$_1$ state. The progression is therefore assigned as 1$^n_0$.

The ground state of SiC$_2$ has a highly anharmonic potential energy surface, in contrast to the first excited state, which is quite harmonic at the triangular geometry \citep{butenhoff91a}. Rather harmonic potentials are also observed for the S$_2$ and S$_5$ states as demonstrated by the almost evenly spaced 1$^n_0$ and 2$^n_0$ bands. Spectral scans at higher resolution could provide a further insight to the assignments.


\subsection{Si$_2$C} \label{sec_Si2C}
The only experimental data on Si$_2$C are IR spectra of mixed carbon-silicon vapors trapped in Ar matrices \citep{kafafi83,presillamarquez91}. Absorptions at 839.5 cm$^{-1}$ and 1188.4  cm$^{-1}$ were ascribed to the symmetric Si--Si stretching fundamental $\nu_1$(a$_1$) and the antisymmetric Si--C stretching vibration $\nu_3$(b$_2$), respectively. The third mode $\nu_2$(a$_1$), a symmetric bending, was inferred indirectly from an observed combination band to be at 166.4 cm$^{-1}$.

\begin{figure}\begin{center}
\epsscale{1.0} \plotone{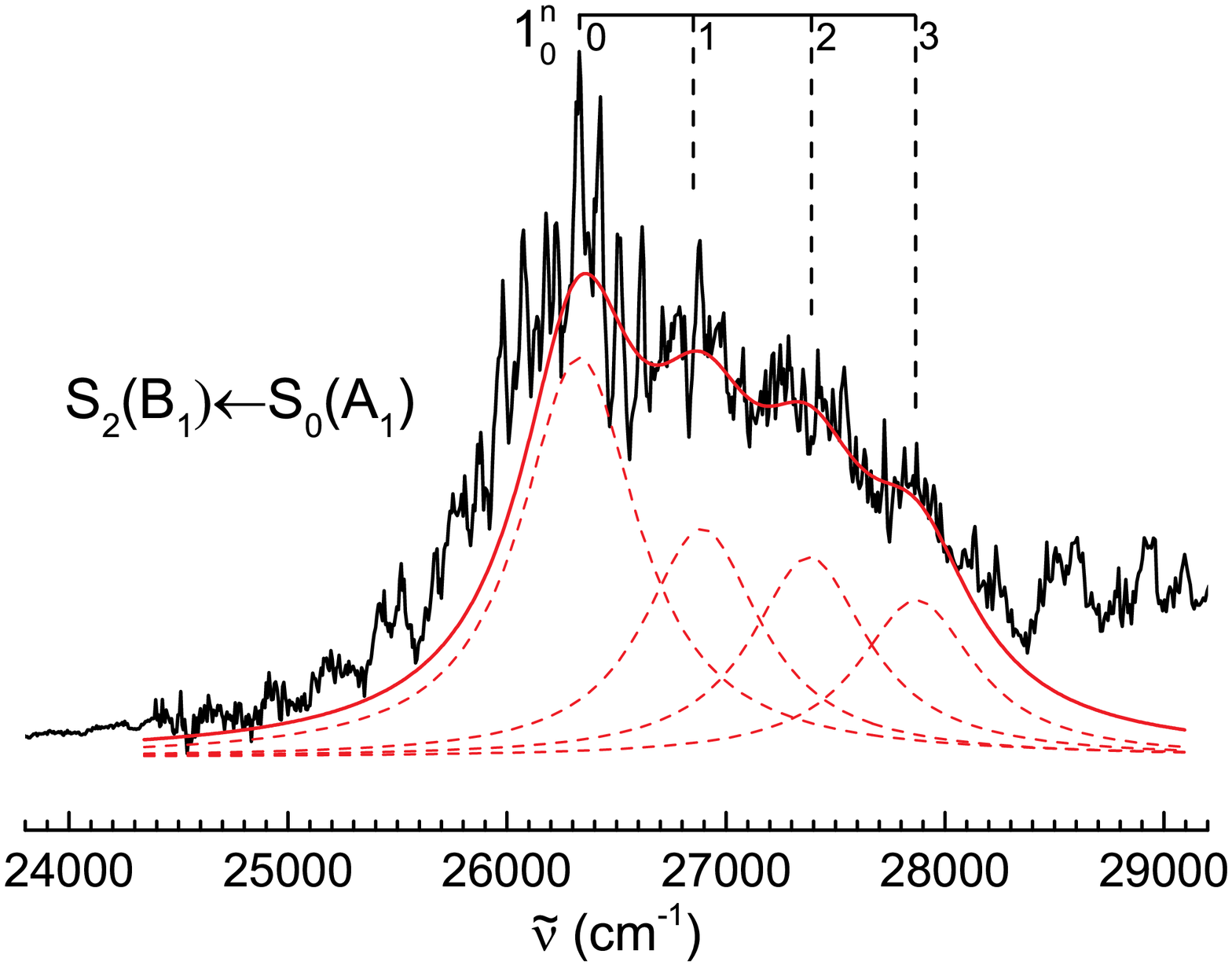} \caption{Vibrational progression in the S$_2$$\leftarrow$S$_0$ spectrum of Si$_2$C.} \label{fig_Si2C}
\end{center}\end{figure}

Three broad absorption profiles with onsets at about 395, 350, and 300 nm are apparent in the Si$_2$C spectrum (Fig. \ref{fig_overview}). According to the TDDFT calculations, they can be caused by several electronic transitions with notable oscillator strengths, probably overlapping each other (from S$_2$ up to S$_8$; see Table \ref{tab_results}). The first singlet transition to an A$_2$ state is symmetry-forbidden.

The profile extending from 395 to 350 nm, ascribed to S$_2$(B$_1$)$\leftarrow$S$_0$(A$_1$), features a partially resolved vibrational progression with a slightly decreasing band separation of ($550\pm50$) cm$^{-1}$ to ($490\pm50$) cm$^{-1}$ (Fig. \ref{fig_Si2C}). The excitation of $\nu_3$(b$_2$) in S$_2$$\leftarrow$S$_0$ is dipole forbidden and the energy of $\nu_2$(a$_1$) is considerably lower. The progression can therefore be assigned to 1$^n_0$. Hot bands on the red side of the profile are vaguely perceptible.

Calculations suggest that the molecule is linear (D$_{\infty h}$) in the S$_2$ state. The symmetric Si--Si stretching vibration $\nu_1$(a$_1$) is now better described as a symmetric Si--C--Si stretch $\nu_1$($\sigma_g$). The computed energy of this mode is 510 cm$^{-1}$ (B3LYP/aug-cc-pVQZ), in quite good agreement with the observed 1$^n_0$ progression. The symmetric bending mode $\nu_2$(a$_1$) becomes $\nu_2$($\pi_g$), calculated at 193 cm$^{-1}$. Because of the geometry change to the linear form, this mode should be considerably populated upon S$_2$$\leftarrow$S$_0$ excitation. Finally, the antisymmetric Si--C--Si stretching vibration $\nu_3$($\sigma_u$) is predicted at 999 cm$^{-1}$ in the excited state.

Line widths in the S$_2$$\leftarrow$S$_0$ bands could be as broad as 500 cm$^{-1}$, corresponding to a natural lifetime on the order of 10 fs. In view of the comparatively large and irregular jitter of the ionization laser (40 ns, pulse width of about 5 ns), an excited state with such a short lifetime should have been only hardly detectable unless a relaxation to another long-lived state above the ground state takes place. Other broadening mechanisms caused by the energetic overlap of several transitions might have to be considered as well.

The appearance of the Si$_2$C electronic spectrum displaying only broad bands in the UV impedes a detection by UV-visible astronomy. An observation is more likely in the radio range (calculated dipole moment of 0.71 Debye).

\subsection{Si$_2$C$_2$} \label{sec_Si2C2}
So far, Si$_2$C$_2$ was only observed by IR spectroscopy in Ar matrices. Two vibrational modes belonging to the rhombic isomer (D$_{2h}$) were identified \citep{presillamarquez95}. At the B3LYP/aug-cc-pVQZ level of theory, the linear D$_{\infty h}$ structure is essentially isoenergetic (only 0.01 eV higher than D$_{2h}$). The electronic spectrum presented in Fig. \ref{fig_overview} exhibits two broad features in the UV and four narrow bands around 515 nm. The first UV feature extends from about 365 to 320 nm. Two maxima separated by ca. 890 cm$^{-1}$ can be recognized. Substantial bandwidths prevent a further analysis. The second UV feature between 295 and 260 nm consists of several broad bands with irregular separation. Upon comparison with TDDFT results, it is found that both UV systems can be caused by the D$_{2h}$ as well as the D$_{\infty h}$ isomers (see Table \ref{tab_results}). An overlap of several electronic transitions, especially within the second absorption system, is possible. 

\begin{figure}\begin{center}
\epsscale{1.0} \plotone{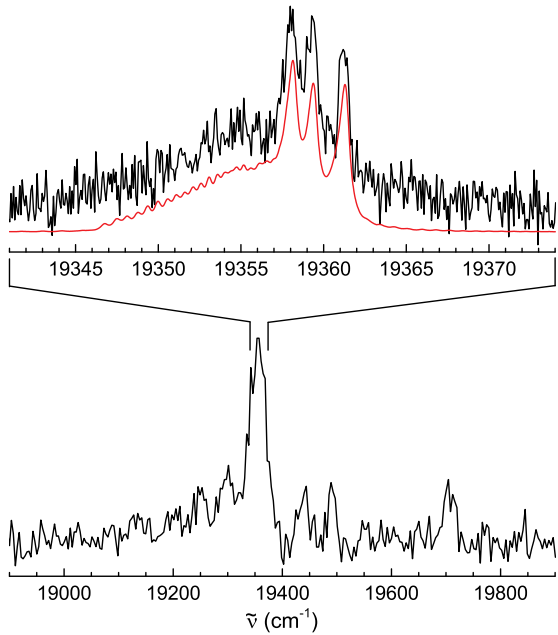} \caption{Detailed view of the green absorption feature of Si$_2$C$_2$. Bottom: low resolution (0.1 nm). Top: high-resolution scan (0.002 nm; black curve) and rotational profile fit assuming the D$_{\infty h}$ isomer ($T_{\text{rot}}$ = 80 K; red trace).} \label{fig_HR-Si2C2}
\end{center}\end{figure}

\begin{table}
 \caption{Spectroscopic constants (in cm$^{-1}$) of linear Si$_2$C$_2$ obtained from the 0$_0^0$ band of $\widetilde{C}^3\Sigma_u\leftarrow \widetilde{X}^3\Sigma_g$.}
\begin{tabular}{|r|cc|} \hline
										& calculation$^{\text{a}}$		& fit	\\ \hline \hline
origin									& 20567					& 19358.6$\pm$0.4	\\
B$_{\text{v}}'$								& 0.0519					& 				\\
B$_{\text{v}}''$ 							& 0.0496					& 				\\
spin-spin $\lambda'$					& 							& -0.80$\pm$0.20	\\
		$\lambda''$					& 							& 1.76$\pm$0.20	\\
spin-rotation $\gamma'$ 				& 							& -0.100$\pm$0.010	\\
			$\gamma''$				& 							& -0.115$\pm$0.010	\\ 
centrifugal d. D$_{\text{v}}'$				& $0.241\times10^{-8}$	&				\\ 
					D$_{\text{v}}''$		& $0.256\times10^{-8}$	& 				\\ \hline
\end{tabular}\\
\begin{tabular}{l}
\footnotesize $^a$ B3LYP/aug-cc-pVQZ \\
\end{tabular}
\label{tab_Si2C2}
\end{table}

A detailed view of the green bands and a high-resolution scan of the strongest one at 516.4 nm are presented in Fig. \ref{fig_HR-Si2C2}. The rhombic isomer is predicted to have two symmetry-forbidden transitions at 492 and 468 nm. The observed bands could arise from vibrational excitations with \textit{ungerade} symmetry. However, the calculated vertical ionization energy (IE) of the D$_{2h}$ structure is 9.01 eV (adiabatic: 8.60 eV), 0.8 eV (0.4 eV) higher than the sum of both the photon energies provided at these wavelengths, making this assignment rather unlikely. The predicted IE of the linear molecule, on the other hand, is sufficiently low for a two-color absorption (vertical: 7.58 eV, adiabatic: 7.48 eV). The first allowed transition $\widetilde{C}^3\Sigma_u\leftarrow \widetilde{X}^3\Sigma_g$ is calculated at 465 nm (adiabatic: 486 nm) and has a fairly high oscillator strength $f=0.25$.

The high-resolution profile of the 516.4 nm band displays a peculiar shape. It consists of a broad plateau superimposed by three narrower features. Rotational lines are not resolved. This profile can be modeled by a linear molecule Hamiltonian containing spin-spin interactions. Fig. \ref{fig_HR-Si2C2} displays the rotational profile fit, assuming a $^3\Sigma_u$$\leftarrow$$^3\Sigma_g$ transition. The rotational constants in the ground and excited states B$_{\text{v}}'$ and B$_{\text{v}}''$ were calculated by DFT and TDDFT, respectively, and kept fixed during the fitting procedure. Centrifugal distortion (D$_{\text{v}}'$, D$_{\text{v}}''$) was also considered, but it has an almost negligible influence on the overall profile. The spin-spin ($\lambda'$, $\lambda''$) and spin-rotation ($\gamma'$, $\gamma''$) coupling constants were allowed to float. All calculated and fitted values are summarized in Table \ref{tab_Si2C2}. The rotational temperature is estimated at $80\pm40$ K. Lorentzian line widths are 0.5 cm$^{-1}$. It is found that the spin-spin coupling is most important to simulate the separation of the three narrower features correctly. In particular, $\lambda$ has to be negative in the ground and positive in the excited state. The difference can be explained qualitatively. Spin-spin interaction is usually large when electronic $\Pi$ states are nearby \citep{herzberg50}. Three $^3\Pi$ states (with low $f$-values from $\widetilde{X}^3\Sigma_g$) are predicted in close vicinity to $\widetilde{C}^3\Sigma_u$.

In addition to the strong band at 516.4 nm (19359 cm$^{-1}$), three weaker bands at 19442, 19490, and 19703 cm$^{-1}$ are recognizable in the low-resolution spectrum. Franck-Condon simulations predict a vibrational progression (including hot bands), which differs from the observed one. A vibronic interaction with one of the nearby $\Pi$ states is possible. Additional high-resolution and in-depth theoretical studies are needed for further clarification.

Linear and rhombic Si$_2$C$_2$ molecules have no permanent dipole moment. An astronomical detection of the D$_{\infty h}$ isomer is now possible in the visible with the data presented here. The 516.4\,nm band is close to three recently discovered DIBs between 517 and 518 nm \citep{hobbs08}. The closest of these three, however, is 0.6 nm away.

\subsection{Si$_2$C$_3$} \label{sec_Si2C3}
The linear Si$\dbond$C$\dbond$C$\dbond$C$\dbond$Si molecule has been studied by anion photoelectron spectroscopy and IR spectroscopy in matrix and gas phase \citep{orden94,presillamarquez94,nakajima95,duan02,thorwirth11}. The electronic spectrum as presented in Fig. \ref{fig_overview} is dominated by two absorption systems starting at about 335 and 300 nm. They can be assigned to the first allowed transitions, $\tilde{F}^1\Sigma_u\leftarrow \tilde{X}^1\Sigma_g$ ($f=1.77$) and $\tilde{G}^1\Pi_u\leftarrow \tilde{X}^1\Sigma_g$ ($f=0.10$), calculated at 292 and 279 nm.

\begin{figure}\begin{center}
\epsscale{1.0} \plotone{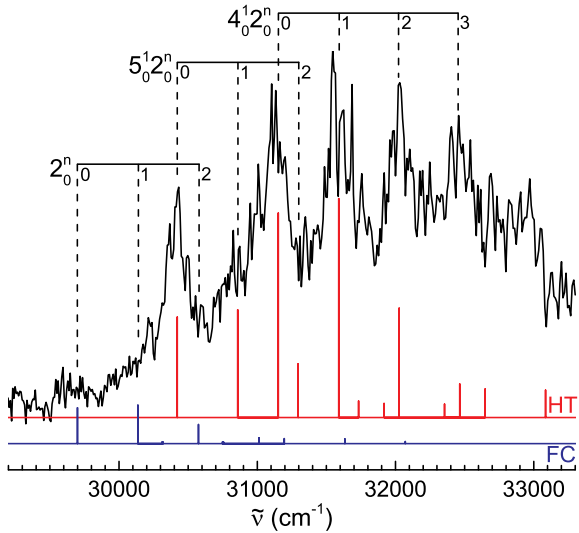} \caption{$\widetilde{F}^1\Sigma_u\leftarrow \widetilde{X}^1\Sigma_g$ transition of Si$_2$C$_3$ compared to the Franck-Condon and Herzberg-Teller simulation ($T_{\text{vib}} = 0 \text{ K}$).} \label{fig_Si2C3}
\end{center}\end{figure}

The $\widetilde{F}^1\Sigma_u\leftarrow \widetilde{X}^1\Sigma_g$ transition displays several broad bands of about 150 cm$^{-1}$ widths (Fig. \ref{fig_Si2C3}). The vibrational structure can hardly be rationalized by a Franck-Condon approach. The strongest bands are likely caused by $5_0^1 2_0^n$ and $4_0^1 2_0^n$ progressions. The corresponding modes are computed at $\nu_2''(\sigma_g)=437\text{ cm}^{-1}$, $\nu_4''(\sigma_u)=1453\text{ cm}^{-1}$, and $\nu_5''(\sigma_u)=721\text{ cm}^{-1}$ in the excited state. In the ground state, they are $\nu_2'(\sigma_g)=471\text{ cm}^{-1}$, $\nu_4'(\sigma_u)=2036\text{ cm}^{-1}$, and $\nu_5'(\sigma_u)=920\text{ cm}^{-1}$. Assignments are based on a Franck-Condon-Herzberg-Teller simulation with the Gaussian09 software, which accounts for a linear variation of the dipole moment during the transition \citep{frisch13}. An excess of absorption is observed on the blue side of the spectrum ($>$32000 cm$^{-1}$). A mixing with the nearby $\widetilde{G}^1\Pi_u$ state is possible. Lacking a permanent dipole moment and strong visible absorptions, this molecule might prove difficult to detect in space.

\subsection{Si$_3$C} \label{sec_Si3C}
\begin{figure}\begin{center}
\epsscale{1.0} \plotone{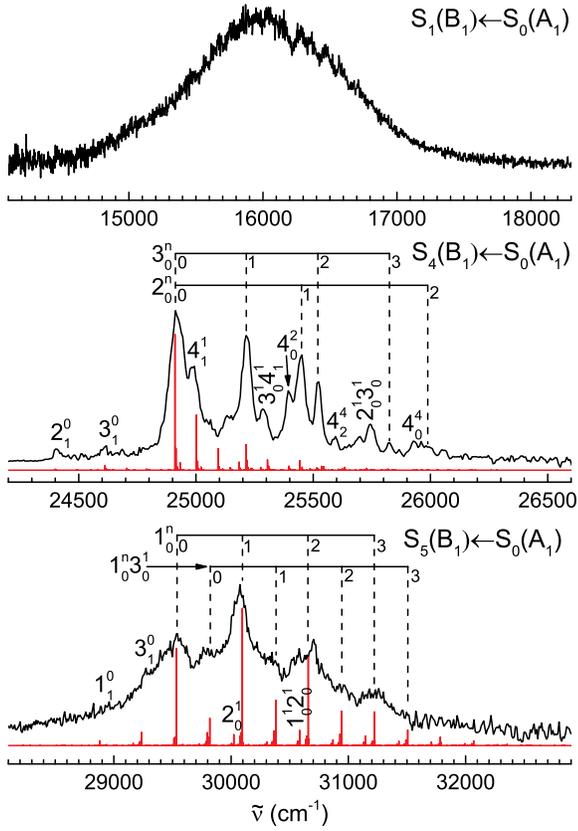} \caption{Electronic transitions of Si$_3$C (black) in comparison to vibrational patterns calculated with the Franck-Condon approach ($T_{\text{vib}}$ = 300\,K; red).} \label{fig_Si3C}
\end{center}\end{figure}

Photodissociation experiments suggest that Si$_3$C is a particularly stable cluster \citep{ticknor05}. Infrared spectroscopic studies in Ar matrices found excellent agreement with calculated vibrational frequencies of the rhomboidal isomer \citep{presillamarquez92}. The band system around 400 nm (Fig. \ref{fig_overview}) has already been observed by R2C2PI and analyzed \citep{stanton05}. It was assigned as $\widetilde{C}^1\text{B}_1\leftarrow\widetilde{X}^1\text{A}_1$ based on equation-of-motion coupled-cluster calculations. At the B3LYP/aug-cc-pVQZ level of theory, this transition appears as $\widetilde{D}^1\text{B}_1\leftarrow\widetilde{X}^1\text{A}_1$ (or S$_4$(B$_1$)$\leftarrow$S$_0$(A$_1$)) and is labeled accordingly here. In addition, the S$_1$(B$_1$)$\leftarrow$S$_0$(A$_1$) and S$_5$(B$_1$)$\leftarrow$S$_0$(A$_1$) transitions were observed around 630 and 330 nm. Fig. \ref{fig_Si3C} is a detailed view of the three absorption systems along with Franck-Condon simulations of the vibrational progressions. Band assignments are listed in Table \ref{tab_Si3C}. An extensive spectroscopic discussion is presented in another paper \citep{reilly14}.

\begin{table}
 \caption{Si$_3$C electronic absorptions (Fig. \ref{fig_Si3C}).}
\begin{tabular}{|lccc|} \hline
     			  					& 						& $\lambda$ 			& $\Delta \nu$		\\
     		  						& band					& (nm)					& (cm$^{-1}$)$^{\text{a}}$	\\ \hline \hline
S$_1$(B$_1$)					& -- 					& 625					& --				\\  \hline
S$_4$(B$_1$)					& 2$^0_1$				& 409.7				& -509				\\
								& 3$^0_1$				& 406.1				& -296				\\
								& 0$^0_0$				& 401.3				& 0					\\
								& 4$^1_1$				& 400.0				& 80				\\
								& -- 					& 397.8					& 219				\\
								& 3$^1_0$				& 396.5				& 303				\\
								& 3$^1_0$4$^1_1$	& 395.4				& 373				\\
								& 4$^2_0$				& 393.7				& 480				\\
								& 2$^1_0$				& 392.8				& 537				\\
								& 3$^2_0$				& 391.7				& 609				\\
								& 4$^4_2$				& 390.6				& 681				\\
								& -- 					& 389.0				& 786				\\
								& 2$^1_0$3$^1_0$	& 388.3				& 831				\\
								& 3$^3_0$				& 387.1				& 911				\\
								& 4$^4_0$				& 385.6				& 1016				\\  \hline
S$_5$(B$_1$)					& 3$^0_1$				& 341.5				& -265				\\
								& 0$^0_0$				& 338.4				& 0					\\
								& 3$^1_0$				& 335.4				& 266				\\
								& 1$^1_0$				& 332.4				& 538				\\
								& 1$^1_0$3$^1_0$	& 329.4				& 812				\\
								& 1$^1_0$2$^1_0$	& 326.9					& 1042				\\
								& 1$^2_0$				& 325.6				& 1163				\\
								& 1$^2_0$3$^1_0$	& 322.9				& 1424				\\
								& 1$^3_0$				& 320.1				& 1690				\\  \hline
\end{tabular}\\
\begin{tabular}{l}
\footnotesize $^{\text{a}}$$\pm$ 10 cm$^{-1}$ \\
\end{tabular}
\label{tab_Si3C}
\end{table}

The S$_1$$\leftarrow$S$_0$ transition in the red displays a broad (1336 cm$^{-1}$) profile of almost perfect Gaussian shape. According to the calculations, S$_1$ is a transition state that leads to a different isomer. The fully optimized geometry differs substantially from that of the ground state. It has a C$_{2v}$ symmetry with a triangular Si$_2$C ring and an exocyclic Si atom \citep[see also ][]{stanton05}. Because of the huge geometry difference, the Franck-Condon approach is actually unsuitable for the prediction of the pattern. Nevertheless, multiple excitations of vibrational modes during the electronic transition can be expected and, in addition to lifetime broadening of the individual bands, likely lead to the observed unstructured profile.

The optimized geometry of S$_4$ is similar to the ground state structure. Thus the most intense feature of S$_4$$\leftarrow$S$_0$ is the origin band at 401.3 nm. Excitations involving the symmetric deformation vibrations $\nu_2$(a$_1$) and $\nu_3$(a$_1$) dominate the rest of the spectrum. Weaker features are due to the low-frequency $\nu_4$(b$_1$) out-of-plane umbrella mode, hot bands, as well as combination bands. It was suggested that a fast ($\sim$ps) internal conversion to a long-lived triplet state causes the observed band widths, which are about 10 to 20 cm$^{-1}$ \citep{stanton05}.

The band widths in the S$_5$$\leftarrow$S$_0$ electronic spectrum are on the order of 100 cm$^{-1}$ indicating an even shorter lifetime. The origin band is located at 338.5 nm. Excitations of the symmetric breathing vibration $\nu_1$(a$_1$) are mainly responsible for the observed progression. The 1$_0^1$ band is the strongest feature of this transition.

An optical detection of Si$_3$C in space might be accomplished by searching for its visible and near-UV transitions. A detection of pure rotational lines is less likely. The calculated dipole moment is only 0.09 Debye.

\subsection{Si$_3$C$_2$} \label{sec_Si3C2}
Three vibrational modes of Si$_3$C$_2$ have been identified in Ar matrices, confirming the pentagonal C$_{2v}$ structure of the ground state \citep{presillamarquez96}. The observed frequencies, $\nu_2$(a$_1$) = 681.1 cm$^{-1}$, $\nu_7$(b$_2$) = 956.7 cm$^{-1}$, and $\nu_8$(b$_2$) = 597.8 cm$^{-1}$, agree well with calculated values applying the B3LYP/aug-cc-pVQZ method (683, 987 and 599 cm$^{-1}$). The other totally symmetric modes are computed as $\nu_1$(a$_1$) = 1501 cm$^{-1}$, $\nu_3$(a$_1$) = 468 cm$^{-1}$, and $\nu_4$(a$_1$) = 168 cm$^{-1}$.

\begin{figure}\begin{center}
\epsscale{1.0} \plotone{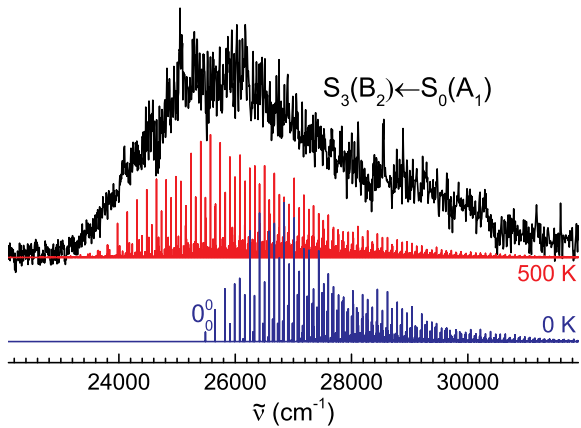} \caption{S$_3$$\leftarrow$S$_0$ spectrum of Si$_3$C$_2$ and Franck-Condon simulation at two different temperatures.} \label{fig_Si3C2}
\end{center}\end{figure}

The main feature of the electronic spectrum is a broad ($\sim$ 4000 cm$^{-1}$) structure, which can be assigned to the strong S$_3$(B$_2$)$\leftarrow$S$_0$(A$_1$) transition. The theoretical oscillator strength is $f=0.12$. The first allowed transition, S$_1$(B$_2$)$\leftarrow$S$_0$(A$_1$), is comparatively weak ($f=0.004$) and calculated at 722 nm, outside of the scanning range. The second transition, S$_2$(A$_2$)$\leftarrow$S$_0$(A$_1$), at 597 nm is symmetry-forbidden. Another broad and unstructured absorption system starting at ca. 290 nm can probably be attributed to the S$_8$(B$_1$)$\leftarrow$S$_0$(A$_1$) or S$_9$(A$_1$)$\leftarrow$S$_0$(A$_1$) transition.

A considerable geometry change is predicted upon S$_3$$\leftarrow$S$_0$ excitation. The excited state has C$_s$ symmetry. Excitations of all four a$_1$ vibrational modes can be expected, which is confirmed by a Franck-Condon simulation (Fig. \ref{fig_Si3C2}). For example, excitations of six and more quanta of the lowest-frequency mode a$_1$ contribute distinct intensity to the vibrational progression. Hot bands are important at vibrational temperatures typical for molecular beam conditions ($T_{\text{vib}}$ = 200--1000 K). The onset of the absorption profile depends heavily on $T_{\text{vib}}$ (see Fig. \ref{fig_Si3C2}). The exact position of the origin band is unknown. No clear vibrational structure is resolved in the experimental spectrum due to spectral congestion and lifetime broadening. Band widths are probably around 100 cm$^{-1}$, which is comparable to the near-UV transitions of Si$_3$C. The weak S$_5$(B$_1$)$\leftarrow$S$_0$(A$_1$) transition ($f=0.005$) might contribute at 353 nm (28300 cm$^{-1}$), where a faint shoulder is recognizable.

The calculated dipole moment of Si$_3$C$_2$ is 1.0 Debye. Considering the broad absorption in the UV, a search for this molecule might prove more successful in the radio range.

\subsection{SiC$_6$H$_4$} \label{sec_SiC6H4}
\begin{figure}\begin{center}
\epsscale{1.0} \plotone{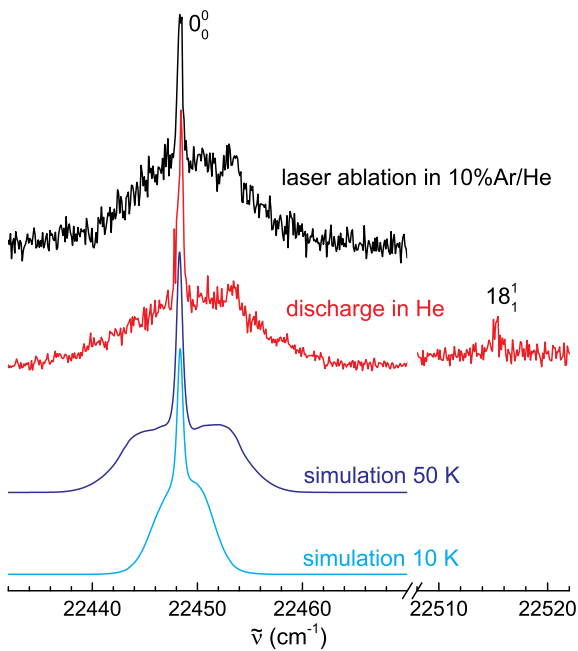} \caption{High-resolution (0.002 nm) scans of the mass 104 spectrum compared to calculated absorption profiles of the S$_1$(B$_1$)$\leftarrow$S$_0$(A$_1$) origin band of SiC$_6$H$_4$.} \label{fig_HR104}
\end{center}\end{figure}

On mass 104, a simple spectrum was measured consisting of two bands, a strong one at 445.3 nm and one at 444.0 nm, which is about five times weaker (Figs. \ref{fig_overview} and \ref{fig_HR104}). The plotted spectra were obtained by using the OPL at 215 nm (5.8 eV) as an ionization laser. Scans with the ArF excimer laser (6.4 eV) revealed no additional bands further to the red. Considering the other species emerging from the ablation source, it is tempting to assign the spectrum to the linear Si--C$_4$--Si triplet, the lowest energy isomer of Si$_2$C$_4$. The benzene-like SiC$_6$H$_4$, however, has the same mass and its calculated spectral properties are in better agreement with the measurements as outlined in the following.

\begin{table}
 \caption{Spectroscopic constants (in cm$^{-1}$) of SiC$_6$H$_4$ in S$_0$ and S$_1$.}
\begin{tabular}{|c|cc|} \hline
										& calculation$^{\text{a}}$& fit	$^{\text{b}}$	\\ \hline \hline
origin									& 21514					& 22448.4$\pm$0.4	\\
A$'$									& 0.1903					& 						\\
B$'$ 									& 0.0666					& 						\\
C$'$									& 0.0494					& 						\\
A$''$									& 0.1883					& 0.190$\pm$0.005	\\
B$''$ 									& 0.0674					& 0.067$\pm$0.005	\\
C$''$									& 0.0496					& 0.049$\pm$0.002	\\
$\nu_{18}''$(b$_1$)-$\nu_{18}'$(b$_1$)	& 80.8						& 66.9$\pm$0.2		\\ \hline
\end{tabular}\\
\begin{tabular}{l}
\footnotesize $^a$ B3LYP/aug-cc-pVQZ \\
\footnotesize $^b$ obtained by fitting the origin band of S$_1$(B$_1$)$\leftarrow$S$_0$(A$_1$) \\
\end{tabular}
\label{tab_SiC6H4}
\end{table}

The first allowed transition of Si$_2$C$_4$ is predicted at 575 nm and is very strong ($f=0.15$), but 0.63\,eV away from the observed band position. This transition has $\Sigma_u\leftarrow\Sigma_g$ symmetry. Its origin band would feature neither the strong Q branch nor the spin-orbit split necessary to explain the measured band profile and the presence of the weak 444.0 nm band. Excited state vibrations were calculated and compared to those in the ground state. A vibration or hot band of Si$_2$C$_4$ cannot explain the weak band either. The computed S$_1$(B$_1$)$\leftarrow$S$_0$(A$_1$) transition of SiC$_6$H$_4$ ($f=0.012$), on the other hand, appears at 410 nm, within 0.24 eV of the measured 445.3 nm band. The calculated position of the origin band is even closer (465 nm, 0.11 eV away). The next allowed ($f=0.013$) transition S$_3$(B$_2$)$\leftarrow$S$_0$(A$_1$) is predicted at 284 nm where the laser power of the OPL is about ten times lower than at 445 nm, but the real transition might be located further in the UV, outside of the scan range, and therefore it could have escaped detection. 

The same spectrum was also observed by discharging phenylsilane (SiC$_6$H$_8$) evaporated at room temperature in He atmosphere (5 bar back pressure on pulse valve). Fragments of the parent molecule were the main species that appeared in the mass spectra, i.e., SiC$_6$H$_4$ should have been much more abundant than Si$_2$C$_4$ in these experiments, confirming the above assignment. A simulation of the band profile based on calculated SiC$_6$H$_4$ geometries in the ground and excited states is in good agreement with the observed profile (see Fig. \ref{fig_HR104}). The band at 444.0 nm can be explained by the $18^1_1$ hot band.  The low-frequency mode $\nu_{18}$(b$_1$) is calculated at 153 cm$^{-1}$ in S$_0$ and 234 cm$^{-1}$ in S$_1$. Table \ref{tab_SiC6H4} lists the theoretical and deduced spectroscopic constants. The hot band intensity could be lowered by adding 10\% of Ar to the supersonic expansion. This decreased the vibrational temperature from ($150 \pm 10$) K to ($120 \pm 10$) K and the rotational temperature from ($70 \pm 10$) K to ($50 \pm 5$) K, as concluded from a Franck-Condon analysis and rotational profile fits. The overall calculated vibrational pattern does not agree with the observation of only two bands. Predissociation can be excluded. The minimum energy to remove a hydrogen (or silicon) is calculated at 4.5 eV (4.7 eV). An interaction with the nearby dark state S$_2$(A$_2$) is possible.

A two-color photoionization threshold scan was conducted to determine the IE of the absorbing species. The band intensity was monitored while keeping the wavelength of the first OPL at 445.3 nm (2.78 eV) and scanning the wavelength of the ionization laser, a second OPL, between 210 and 224\,nm (5.90 -- 5.54 eV). The IE of ($8.395 \pm 0.015$) eV thus obtained is in agreement with theoretical predictions for SiC$_6$H$_4$ (vertical IE = 8.47 eV, adiabatic IE = 8.24 eV). At the same time, the calculated value of Si$_2$C$_4$ differs by more than 1\,eV (vertical IE = 7.29 eV). By measuring the 445.3 nm band intensity while varying the delay between scan and ionization laser the excited state lifetime was determined to be $\tau = (8 \pm 1)$\,ns.


This is the first spectroscopic study of SiC$_6$H$_4$. Aside from its absorptions in the visible, it has a strong dipole moment, calculated at 1.65 Debye. Therefore, SiC$_6$H$_4$ could be easier to find in space than benzene, which was discovered only by infrared observations of a protoplanetary nebula \citep{cernicharo01}.

\section{Summary} \label{sec_summary}
Several silicon-carbon molecules have been prepared by laser ablation and studied by R2C2PI spectroscopy between 250 and 710 nm. New bands and transitions were found for SiC$_2$ and Si$_3$C. The molecules Si$_2$C, Si$_2$C$_2$, Si$_2$C$_3$, and Si$_3$C$_2$ were investigated for the first time by electronic spectroscopy. This is also the first experimental study of the linear Si$_2$C$_2$ isomer and the benzene-like SiC$_6$H$_4$ radical. Aside from SiC$_2$, which has already been found in space, linear Si$_2$C$_2$, Si$_3$C, and SiC$_6$H$_4$ are interesting target molecules for astronomical searches in the visible due to their spectral properties. A band at 516.4 nm with a rather unique shape due to spin-spin splitting was found for linear Si$_2$C$_2$. The calculated oscillator strength of the corresponding transition is fairly high ($f = 0.25$). SiC$_6$H$_4$ could be easier to detect than benzene because of its transition at 445.3\,nm and its strong dipole moment. The simplicity of the electronic spectrum could also hint at the DIB problem. Like SiC$_6$H$_4$, the carriers of these enigmatic bands probably do not feature complicated vibrational progressions and strong UV bands as most DIBs do not correlate with each other and none have been found below 400\,nm.

\acknowledgments
This work has been funded by the Swiss National Science Foundation (Project 200020-140316/1).

\end{document}